\documentclass[twocolumn,amsfonts,amsmath,amssymb,final,aps,prd,footinbib,longbibliography,nofootinbib,preprintnumbers,10pt]{revtex4-2}
\usepackage[dvipsnames]{xcolor}
\usepackage{dcolumn}
\usepackage[colorlinks=true,allcolors=Blue]{hyperref}
\usepackage{upgreek,graphicx,manfnt,pifont,colonequals,bbm,stmaryrd}
\usepackage[shortlabels]{enumitem}
\usepackage{amsthm}

\def\be{\begin{eqnarray}}
	\def\ee{\end{eqnarray}}
\def\bea{\begin{eqnarray}}
	\def\eea{\end{eqnarray}}

\newcommand{\Tr}{\mbox{Tr}}

\def\S{{\mathbb S}}

\def\U{\text{U}}
\def\SU{\text{SU}}
\def\USp{\text{USp}}
\def\SO{\text{SO}}
\def\E{\text{E}}

\begin{document}
\preprint{YITP-SB-2025-12}
	\title{Generalized Schur partition functions and RG flows}
	
	\medskip
	
	\author{Anirudh Deb}

	\affiliation{C. N. Yang Institute for Theoretical Physics, Stony Brook University, Stony Brook, NY 11794-3840, USA}

	\author{Shlomo S. Razamat}
	\affiliation{Department of Physics, Technion, Haifa 32000, Israel}

	\begin{abstract}
		\noindent We revisit a double-scaled limit of the superconformal index of ${\cal N}=2$ superconformal field theories (SCFTs) which generalizes the Schur index. The resulting partition function, $\hat {\cal Z}(q,\alpha)$, has a standard $q$-expansion with coefficients depending on a continuous parameter $\alpha$. The Schur index is a special case with $\alpha=1$. Through explicit computations we argue that this partition function is an invariant of certain mass deformations and vacuum expectation value (vev) deformations of the SCFT. In particular, two SCFTs residing in different corners of the same Coulomb branch, satisfying certain restrictive conditions,   have the same partition function with a non-trivial map of the parameters, $\hat {\cal Z}_1(q,\alpha_1)=\hat {\cal Z}_2(q,\alpha_2(\alpha_1))$. For example, we show that the Schur index of {\it all}  the SCFTs in the Deligne-Cvitanovi\'c series is given by special values of $\alpha$ of the partition function of the $\SU(2)$ $N_f=4$ ${\cal N}=2$ SQCD. 
		
	\end{abstract}	
	\maketitle
	\noindent{{\bf Introduction}: }
         The dynamics of quantum field theories with eight supercharges is a focal point of 
	numerous research directions in quantum field theory and mathematical physics.
	An example of such a convergence of naively disparate ideas is given by the {\it Schur index} \cite{Gadde:2011ik,Gadde:2011uv}. On the one hand, the Schur index is a special case of the supersymmetric index \cite{Kinney:2005ej} of a general ${\cal N}=2$ SCFT counting $\frac14$ BPS operators. However, over the recent years numerous connections between this index and other, sometimes rather different ideas, have been observed/derived. Most notably, the Schur index is the vacuum character of a vertex operator algebra (VOA) that one can associate to an ${\cal N}=2$ SCFT \cite{Beem:2013sza};  it can be thought of as a correlation function in 2d $q$-deformed  YM \cite{Gadde:2011ik}; it is directly related to Kontsevich-Soibelman wall crossing invariants of the Coulomb branch \cite{Cordova:2016uwk}; and more recently connections between Schur index and the double scaled SYK models have been observed \cite{Gaiotto:2024kze}. See also \cite{Gaiotto:2024osr,Ambrosino:2025qpy}. 	
    
The Schur index, naively, is an object which is naturally associated with the Higgs branch of the theory, as we will review soon. However, in a rather surprising manner it also captures some of the properties of the Coulomb branch \cite{Cordova:2015nma,Cordova:2016uwk,Gaiotto:2024ioj}. See also \cite{Kaidi:2022sng,Buican:2024jjl,Fredrickson:2017yka,Pan:2024hcz,Deb:2025cqr} for other attempts to find connections between the Higgs and Coulomb branches. Here we will observe yet another relation between the Schur index and the Coulomb branch. In particular, we will revisit and define a generalized Schur limit of the full supersymmetric index which will depend on two parameters, ${\cal Z}(q,\alpha)$, such that $\alpha=1$  coincides with the Schur index. We will observe that tuning $\alpha$ in certain cases this partition function will also encode Schur indices of theories which can be obtained from the given theory by tuning to  singular loci of it's Coulomb branch.

	\noindent{{\bf Background}:}	
	Let us start by recalling the definition of the full index (a supersymmetric ${\mathbb S}^3\times {\mathbb S}^1$ partition function) of an ${\cal N}=2$ SCFT \cite{Romelsberger:2005eg,Kinney:2005ej,Dolan:2008qi,Gadde:2011uv},
	\be
	{\cal I}=\Tr_{{\mathbb S}^3}\,(-1)^F \left(\frac{qp}t\right)^{-r}p^{j_2-j_1}\,q^{j_1+j_2}\,t^{R}\,\prod_{i=1}^{\text{rank\,} G_F}u_i^{F_i}\,.\;\;
	\ee
	 Here $r$ is the $\U(1)_r$ R-charge, $R$ is the Cartan generator  of the $\SU(2)_R$ R-symmetry, and $j_i$ are the Cartan generators of the $\SU(2)\times \SU(2)$ isometry of ${\mathbb S}^3$. The charges $F_i$ correspond to the Cartan of the flavor symmetry of the theory.  Although our discussion is general, it is useful to understand various statements for a theory defined with a Lagrangian. For that, we remind the reader that the scalars in the chiral field of the vector multiplet, $\Phi$, contribute with weight $\frac{q\,p}t$ to the index; while the scalars of the half hypermultiplets, $Q/\widetilde Q$, have weight $t^{\frac12}$.

	Let us consider turning on a mass term $m\,Q\widetilde Q$, or a vev  $\langle \Tr\Phi^k\rangle\neq 0$. Generically, for the index this will imply that $t^{\frac12}=\sqrt{q\,p}$ in the former case, or $\left(\frac{q\,p}t\right)^k=1$ in the latter: modulo phases that will not be important for us, these are equivalent conditions, $q\,p=t$. Alternatively, setting $q\,p=t$ in the index is equivalent to giving all the hypermultiplets a mass and all the Coulomb branch operators a vev.
	It is easy to see \cite{Gaiotto:2012xa} that in this case the index  will have a pole singularity order of which is the dimension of the Coulomb branch ${\mathfrak r}$ which will multiply $\left((q;q)(p;p)\right)^{{\mathfrak r}}$ where we define $(z;q)=\prod_{i=0}^\infty (1-z\,q^i)$.\footnote{In particular if ${\mathfrak r}=0$, the index will be just $1$.}

	Next, let us remind the definition of the Schur index. If we set $q=t$ the index will depend only on $q$ (and the flavor fugacities),
	\be
	{\cal I}_S=\Tr \,(-1)^F\, q^{R+j_1+j_2}\,\prod_{i=1}^{\text{rank\,} G_F}u_i^{F_i}\,.
	\ee 
	The dependence on $p$ drops out as the charge coupled to it  equals $\{{\cal Q},\, {\cal Q}^\dagger\}$, with ${\cal Q}$ commuting with $R+j_1+j_2$. As Coulomb branch operators of dimension $n$ contribute to the full index as $\left(\frac{q\,p}t\right)^n$, in 
	the  Schur index this contributions becomes $p^n$. Since the Schur limit does not depend on $p$ this contribution has to cancel against another, fermionic, operator. For example, for field $\Phi$ this will be  the gaugino $\lambda$. This is a sense in which Schur index is naturally a Higgs branch object.\footnote{Another, more invariant, way to see this is to note that the Schur index counts operators satisfying the following shortening conditions,
		\begin{equation}
			\begin{split}
				E-(j_1+j_2)-2R=0~,\;\;\;
                \; r+j_1-j_2=0~,
			\end{split}
		\end{equation}
		which contain as a subset, the Higgs branch operators $E=2R, r=0$. More fundamentally, from the point of view of the Higgs branch conjecture \cite{Beem:2017ooy} of the VOA/SCFT correspondence \cite{Beem:2013sza}, the Schur index is more naturally a Higgs branch quantity.}

	Setting $q=t$, the Schur index is consistent with choosing any value for $p$, and in particular setting $p=1$. Although, $q=t$ and $p=1$ satisfies $q\,p=t$, the index does not trivialize as above. We thus observe that we have an {\it order of limits issue}: first setting $q\,p=t$ and then $t=q$  is not the same as first setting $q=t$ and then $p=1$.

	\noindent{{\bf Generalized Schur partition functions}:}	
	This order of limits issue allows us to define an interesting generalized limit of the index which will interpolate between the two extreme cases discussed above.\footnote{Let us stress that the limit we are discussing here was originally defined and studied 
    in \cite{Cecotti:2015lab} with the motivation to relate different {\it integer} values of $\alpha$ to traces of powers of monodromy matrix generalizing the observations of \cite{Cordova:2016uwk,Gaiotto:2024ioj,Cecotti:2010fi}. This line of thought led to various interesting observations, see {\it e.g} \cite{Kim:2024dxu}.
    However, in our work we allow for any non-negative real value of $\alpha$ and do not try to connect with the monodromy matrix.}  We parametrize,
	\be 
	p=1-\epsilon\,,\qquad t=(q\,p)^{1+\frac{\alpha}{\log\,q}\epsilon}\,,
	\ee and take the limit of $\epsilon\to 0$. The case of $\alpha=0$ corresponds to first turning on masses and then taking the Schur limit. The case of $\alpha=1$ corresponds to first taking the Schur limit and then taking $p=1$. This can be clearly seen by computing the (plethystic log of the) contribution of a free hypermultiplet,
	\be
	\frac{t^{\frac12}-\frac{q\,p}{t^{\frac12}}}{(1-q)(1-p)}(z+z^{-1})\;\to\; \alpha\, \frac{q^{\frac12}}{1-q}(z+z^{-1})\,.
	\ee Here $z$ denotes the $U(1)$ flavor fugacity for a full hypermultiplet. The (plethystic log of the) vector field becomes in this limit,
	\be
	&&\left(-\frac{p}{1-p}-\frac{q}{1-q}+\frac{\frac{p\,q}t-t}{(1-p)(1-q)}\right)\,\chi_{adj}({\bf z})\to\\
	&&\;\;\; \;\;\; \left(1-\alpha-\frac{2q}{1-q}\,\alpha\right)\,\chi_{adj}({\bf z})\,.\nonumber
	\ee Here $\chi_{adj}({\bf z})$ is the character of the adjoint representation of the gauge group.
	The first term above contributes a pole singularity of order of the rank of the gauge group. Combining the above for a Lagrangian theory stripping off the pole singularity the limit takes the form,
	\be
	\label{eq:Schurpart}
	{\cal Z}(q,\,\alpha)=(q;q)^{2\,\frak r}\oint d{\bf \zeta}_G \left(\frac{\prod_\beta(1-e^{\beta(\zeta)})(q e^{\beta(\zeta)};q)^2}{\prod_\rho(q^{\frac12}e^{\rho(\zeta)};q)}\right)^\alpha\,.\;\;
	\ee 
	Here $\beta$ are the non-zero roots of the gauge groups and $\rho$ are the weights of the matter representation. For $\alpha=1$ this is the Schur index. For $\alpha=0$ this gives $(q;q)^{2\,\frak r}$, the Schur index of the free $\U(1)$ vectors on a generic locus of the Coulomb branch. For general $\alpha\in {\mathbb R}^+$ we have an interpolating limit that is well defined and that we will study next.
	
	 Let us stress the following point. The limit we consider can be just thought of as a specialization of the full index and therefore makes sense only for conformal theories (as we need both $\SU(2)_R$ and $\U(1)_r$ to define it). In particular, the index of theories related by S-dualities should be the same \cite{Gadde:2009kb}. Moreover, the index should be expandable in terms of sets of wave-functions which will in general depend on $q$ and $\alpha$ and will generalize the Schur polynomials \cite{Gadde:2011uv, Gaiotto:2012xa}.  We comment on this more  in Appendix A.\footnote{The limit we consider is equivalent to taking $p,\, q/t\to 1$ while keeping $q$ generic and keeping track of $\log q/\log t$. If we would also take $q\to 1$, we would obtain the $\S^3$ partition function of the dimensionally reduced theory. Thus, in certain sense, our limit is a $q$-deformation of the ${\mathbb S}^3$ partition function. From this point of view $\alpha$ is proportional to the real mass parameter for $R-r$.
     See {\it e.g.} \cite{Razamat:2014pta,Aharony:2013dha} for some discussions of similar $3d$ limits.}

	\noindent{{\bf Case study of rank $1$}:}	
	Let us discuss the generalized Schur partition function for  arguably the simplest interacting conformal ${\cal N}=2$ SCFT with a manifestly supersymmetric Lagrangian description, the $N_f=4$ $\SU(2)$ SQCD,
	\be
	\label{eq:su2alpha}
	\hat {\cal Z}_{\mathfrak{d}_4}(q,\alpha)=\frac{(q;q)^{2}}{{\mathtt{N}}(\alpha)}\oint\frac{dz}{4\pi i z} \left(\frac{\Delta(z)(q\, z^{\pm2};q)^2}{(q^{\frac12}z^{\pm1};q)^8}\right)^\alpha\,.\quad 
	\ee Here we have defined,
	\be
	&&\Delta(z)=(1-z^2)(1-z^{-2})\,,\quad{\mathtt{N}}(\alpha)=\oint\frac{dz}{4\pi i z}\Delta(z)^\alpha\,.\;\;\;
	\ee We also have defined $\hat {\cal Z}_{\mathfrak{d}_4}(q,\alpha)={\cal Z}_{\mathfrak{d}_4}(q,\alpha)/{\mathtt{N}}(\alpha)$.\footnote{Note that $\mathtt{N}(1)=1$.} One can compute explicitly this normalized partition function in series expansion in $q$. The coefficients will depend on $\alpha$ and will be in general not integer. However, we make the following ``experimental'' observation. Tuning $\alpha$ to certain rational values the expansion coefficients of this normalized partition function are integer, and moreover the partition function can be identified as the Schur index for a different SCFT, properties of which depend on $\alpha$. For example,\footnote{We have computed these expressions expanding the integrand in $q$ and performing the integrals analytically for integer $\alpha$ and numerically for fractional $\alpha$. In the former case it is easy to compute to high orders in $q$, while in the latter due to numerics we could go to low orders only. }
	\be
    &&\hat {\cal Z}_{\mathfrak{d}_4}(q,\frac15)=1+0\,q+q^2+q^3+\cdots={\cal Z}_{\mathfrak{a}_0}(q,1)\,,\\
	&&\hat {\cal Z}_{\mathfrak{d}_4}(q,\frac13)=1+3q+9q^2+19 q^3+\cdots={\cal Z}_{\mathfrak{a}_1}(q,1)\,,\nonumber\\
	&&\hat {\cal Z}_{\mathfrak{d}_4}(q,\frac12)=1+8q+36q^2+128 q^3+\cdots={\cal Z}_{\mathfrak{a}_2}(q,1)\,,\nonumber\\
	&&
    \hat {\cal Z}_{\mathfrak{d}_4}(q,2)=1+78q+2509 q^2\cdots={\cal Z}_{\mathfrak{e}_6}(q,1)\,,\nonumber\\
	&&
    \hat {\cal Z}_{\mathfrak{d}_4}(q,3)=1+133q+7505 q^2\cdots={\cal Z}_{\mathfrak{e}_7}(q,1)\,,\nonumber\\
	&&
    \hat {\cal Z}_{\mathfrak{d}_4}(q,5)=1+248q+27249q^2+\cdots={\cal Z}_{\mathfrak{e}_8}(q,1)\,.\nonumber
	\ee which are precisely the Schur indices of the  the $\mathfrak{a}_0\simeq (A_1,A_2)$, $\mathfrak{a}_1\simeq (A_1,A_3)$, $\mathfrak{a}_2\simeq (A_1, D_4)$ Argyes-Douglas theories \cite{Argyres:1995jj,Argyres:1995xn} and the $\mathfrak{e}_6$,$\mathfrak{e}_7$,$\mathfrak{e}_8$ Minahan-Nemeschansky theories \cite{Minahan:1996fg, Minahan:1996cj},  respectively. More generally we find that,
	\be
	\hat {\cal Z}_{{\mathfrak g}(h_{\mathfrak g}^\vee)}(q,\,\alpha)=\hat {\cal Z}_{\mathfrak{d}_4}(q,\,\frac{h_{\mathfrak g}^\vee}6\,\alpha)\,,
	\ee where $h^\vee_{\mathfrak{g}}$ is the dual-Coxeter number for Lie algebra $\mathfrak{g}$. The above statement is also summarized in Table I.
	%--------------------
	%--------------------
	\begin{table}[t]
		\begin{tabular}{|c|c|c|c|c|c|}
			\hline
			Theory & $h^\vee_{\mathfrak{g}}$& $12c$  & $24a$ & $\Delta$ & $\alpha$\\
			\hline
			$\mathfrak{e}_8$ & $30$ & $62$ & $95$ & $6$  & $5$\\
			\hline
			$\mathfrak{e}_{7\frac{1}{2}}*$ & $24$ & $50$ & $77$ & $5$  & $4$\\ 
			\hline
			$\mathfrak{e}_7$ & $18$ & $38$ & $59$ & $4$ & $3$\\
			\hline
			$\mathfrak{e}_6$ & $12$ &$26$ & $41$& $3$ & $2$\\
			\hline
			$\mathfrak{f}_4*$ & $9$ &$20$ &$32$ & $5/2$ & $3/2$\\
			\hline
			$\mathfrak{d}_4$ & $6$ &$14$ &$23$ & $2$ & $1$\\
			\hline
			$\mathfrak{g}_2*$ & $4$ &$10$ &$17$ & $5/3$ & $2/3$\\
			\hline
			$\mathfrak{a}_2$ & $3$ &$8$ &$14$ & $3/2$ & $1/2$\\
			\hline
			$\mathfrak{a}_1$ & $2$ &$6$ &$11$ & $4/3$ & $1/3$\\
			\hline
			$\mathfrak{a}_{\frac{1}{2}}*$ & $3/2$ &$5$ & $19/2$& $5/4$ &  $1/4$\\
			\hline
			$\mathfrak{a}_0$ & $6/5$ &$22/5$ & $43/5$& $6/5$ &  $1/5$\\
			\hline
		\end{tabular}
		\caption{Values of $\alpha$ for the Deligne-Cvitanovi\'c rank-one series. }
		\label{tab:delrankone}
	\end{table}
	%--------------------
	%-------------------
	Thus the Schur indices of the complete Deligne series of ${\cal N}=2$ SCFTs are encoded in the generalized Schur partition function of the SQCD.\footnote{Note that entries in Table I marked by $*$ do not correspond to known $4d$ SCFTs and do not show up in the rank-one classification of \cite{Argyres:2015ffa,Argyres:2015gha,Argyres:2016xmc,Argyres:2020nrr}. The limit of the index described does not imply existence of such SCFTs, rather the claim is that for the corresponding values of $\alpha$ we recover the vacuum characters of the VOA in the series. For $\alpha=4$, we find the VOA  $(\mathfrak{e}_{7\frac{1}{2}})_{-5}$  \cite{Arakawa:2016hkg,Lee:2023owa}  and $\alpha=\frac{1}{4}$ corresponds to $(\mathfrak{a}_{\frac{1}{2}})_{-\frac{5}{4}}$ \cite{cohen1996computational} (see also \cite{Abhishek:2023wzp,Cho:2024civ}). Moreover, \cite{Cho:2024civ} describe an ${\cal N}=1$ Lagrangian which flows in the IR to an ${\cal N}=2$ theory, with central charges, Coulomb branch dimension, and the Schur index matching the  $(\mathfrak{a}_{\frac{1}{2}})_{-\frac{5}{4}}$. The VOAs with $\mathfrak{f}_4$ and $\mathfrak{g}_2$ do not correspond to vanilla SCFTs but can be related to such by studying dual descriptions of Higgs branches of SCFTs~\cite{Bourget:2020asf}.} The theories in the Deligne-Cvitanovi\'c series are related to each other by Coulomb branch RG flows. Let us comment that for $\alpha>1$ we obtain theories which can be deformed on their Coulomb branch to flow to the SQCD, while for $\alpha<1$ we obtain theories into which SQCD can flow once it's Coulomb branch is explored.\footnote{The Schur indices of the Deligne-Cvitanovi\'c rank-one SCFTs solve a second order differential equation parametrized by a single parameter $h^\vee_{\mathfrak{g}}$  \cite{Beem:2017ooy}. Our results suggest that this can be interpreted as the generalized Schur partition function solving an MLDE with $\alpha$ as a parameter. We have made sporadic checks of this statement.}
  
	It is interesting to note that taking the Schur limit of the integrand for the ${\cal N}=1$ matrix integral expression of the superconformal index \cite{Maruyoshi:2016tqk,Maruyoshi:2016aim} for $\mathfrak{a}_0, \mathfrak{a}_1$ and $\mathfrak{a}_2$ theories leads to the same expression as equation \eqref{eq:Schurpart} for corresponding values of $\alpha$.

	\noindent{{\bf Higher rank and other generalizations}:}
	In the previous section, we have described the generalized Schur limit in the case of the rank-one theories. This can be extended to theories of higher rank. Studying various examples, we propose the following conjectural statement.

    \
    
\noindent {\it 		A necessary condition for the partition function of two theories $\mathcal{T}_1$ and $\mathcal{T}_2$ to be related to each other as follows
		\begin{equation}
			\hat{\mathcal{Z}}_{\mathcal{T}_2}(q,1)=\hat{\mathcal{Z}}_{\mathcal{T}_1}(q,\alpha)
		\end{equation}
		is that their central charges $c^{(1)}$, $c^{(2)}$ and the Coulomb branch scaling dimensions $\Delta_{i=1,2,\dots,\mathfrak{r}}^{(1)}$, $\Delta_{i=1,2,\dots,\mathfrak{r}}^{(2)}$ are related as follows 
		\begin{equation}
			c^{(2)}=c^{(1)}\alpha+\frac{1}{6}(1-\alpha)\mathfrak{r}\,,\;\; \;
			\Delta^{(2)}_i=(\Delta^{(1)}_i-1)\alpha+1,
		\end{equation}
}

\
    
    The matching of the $a-c$ anomaly is natural from the point of view of the index. The leading behavior of a variety of partition functions in the Cardy limit is determined by $a-c$ under some assumptions~\cite{DiPietro:2014bca,ArabiArdehali:2015ybk,Buican:2015ina,ArabiArdehali:2023bpq,Bobev:2015kza}. Thus $a-c$ of the putative theory ${\cal T}_2$ is determined by that of ${\cal T}_1$ and by the value of $\alpha$. In fact looking at the subleading behavior this is also true for the $c$ central charge itself (under certain assumptions) \cite{ArabiArdehali:2015ybk}.\footnote{The matching of $c$ is also expected  from the modular properties of the index \cite{Razamat:2012uv,Beem:2017ooy}.}
    Taking the integrand to the power $\alpha$ multiplies $c$ by $\alpha$, and further there are $2(1-\alpha)\,\mathfrak{r}$ $q$-Pochhamer symbols, $(q;q)^{2(1-\alpha)\mathfrak{r}}$, which are the index of free vector multiplets  contributing $\frac{1}{6}(1-\alpha)\mathfrak{r}$ to $c$.  The combined contribution equals the central charge of the theory obtained in this limit. The above is consistent with the Shapere-Tachikawa relation $4(2a-c)=\sum_{i=1}^{\mathfrak{r}}(2\Delta_i-1)$ \cite{Shapere:2008zf}, and then
		$a^{(2)}=a^{(1)}\alpha+\frac{5}{24}(1-\alpha)\mathfrak{r}$.
    The relation between the Coulomb branch scaling dimensions is an ``experimental'' observation. 
	
	We have tested the above conjecture in various computations. For example, the $\SU(N)$ SQCD with $2N$ flavors and $\USp(2N)$ SQCD with $2N+2$ flavors admit a rather uniform behaviour as tabulated in table \ref{tab:SUN} and \ref{tab:USp2N}. The entries  $\mathfrak{a}_0\simeq (A_1,A_2), \mathfrak{a_1}\simeq (A_1,A_3)\simeq (A_1,D_3), \mathfrak{a_2}\simeq D_2(\SU(3))\simeq (A_1,D_4), \mathfrak{e}_6\simeq R_{2,3}$ in the Deligne-Cvitanovi\'c series  are special cases of tables \ref{tab:SUN} and \ref{tab:USp2N}.\footnote{ We note that $(A_1,D_{2N}), (A_1,A_{2N-1})$ theories are obtained by an $\mathcal{N}=1$ Lagragian obtained from deforming $\mathcal{N}=2$ $\SU(N)$ gauge theory with $2N$ flavors \cite{Maruyoshi:2016aim,Agarwal:2016pjo}. Similarly $(A_1,D_{2N+1}), (A_1,A_{2N})$ are obtained starting from $\mathcal{N}=2$ $\USp(2N)$ $2N+2$ flavors \cite{Maruyoshi:2016aim,Agarwal:2016pjo}.} For $\USp(4)$ SQCD with $6$ flavor, in addition to the entries in tables \ref{tab:USp2N} we also found that the index for $\alpha=3$  coincides with the index of $D_{1}^{20}(\E_8)$ theory.\footnote{In fact the expression with $\alpha=3$ of the $\USp(4)$ model can be thought of as a definition of the Schur index of the $D_{1}^{20}(\E_8)$ theory. Here the independent check we performed is that of anomalies, Coulomb branch dimensions, and the dimension of the flavor symmetry group as read from order $q$ term in the index (and in some cases we could go to higher orders in $q$). 
    For the $\mathfrak{e}_8$ theory the expression using $\alpha=5$ of the $\SU(2)$ SQCD is a novel prediction for a two parameter slice of the index (beyond the Macdonald index which is explicitly known). We would also like to note that similar to the rank-one case, for higher rank cases, the  order of MLDE solved by the Schur indices is the same for theories related by an $\alpha$, in all cases we could check.}

	%--------------------
	%--------------------
	\begin{table}[t]
		\resizebox{\linewidth}{!}{%
			\begin{tabular}{|c|c|c|c|c|c|}
				\hline
				Theory & $12c$ & $24a$  & $\Delta_{i}$ & $\alpha$\\
				\hline
				$R_{2,2N-1}$ &$2\left(4 N^2-N-1\right)$ &$14 N^2-5 N-5$  & $2i+1$ &$2$\\
				\hline
				$\SU(N)$+$2N$F &$4 N^2-2$ &$7 N^2-5$& $i+1$ & $1$\\
				\hline
				$(A_1,D_{2N})$ &$6N-4$ &$2 (6 N-5)$ & $\frac{i}{N}+1$ & $\frac{1}{N}$\\
				\hline
				$(A_1,A_{2N-1})$ &$\frac{6 N^2-2 N-2}{N+1}$ & $\frac{12 N^2-5 N-5}{N+1}$& $\frac{i}{N+1}+1$  & $\frac{1}{N+1}$\\
				\hline
		\end{tabular}}
		\caption{Series obtained from $\SU(N)$ with $2N$ fundamental flavors.}
		\label{tab:SUN}
	\end{table}

	%--------------------
	%--------------------
	\begin{table}[t]
		\resizebox{\linewidth}{!}{%
			\begin{tabular}{|c|c|c|c|c|c|}
				\hline
				Theory & $12c$ & $24a$  & $\Delta_{i}$ &$\alpha$\\
				\hline
				$\USp(2N)$+$(2N+2)$F &$2 N (4 N+3)$ &$N(14 N+9)$& $(2i-1)+1$ & $1$\\
				\hline
				$D_2(SU(2N+1))$ & $4N(N+1)$ & $7N(N+1)$ & $\frac{(2i-1)}{2}+1$ & $\frac{1}{2}$\\
				\hline
				$(A_1,D_{2N+1})$ &$6N$ &$\frac{3 N (8 N+3)}{2 N+1}$ & $\frac{(2i-1)}{2N+1}+1$ & $\frac{1}{2N+1}$\\
				\hline
				$(A_1,A_{2N})$ &$\frac{2 N (6 N+5)}{2 N+3}$ & $\frac{n (24 N+19)}{2 N+3}$& $\frac{(2i-1)}{2N+3}+1$ &$\frac{1}{2N+3}$\\
				\hline
		\end{tabular}}
		\caption{
        Series obtained from $\USp(2N)$ with $2N+2$ fundamental flavors. }
		\label{tab:USp2N}
	\end{table}
	
There are numerous examples which do not satisfy the conditions that we have outlined. The simplest ones are the theories which are related by Coulomb branch deformations to the ${\cal N}=4$ SYM, see {\it e.g.} \cite{Martone:2021ixp,Martone:2021drm}.

	\noindent{{\bf Summary and Discussion}:}
	Supersymmetric partition functions of theories related to each other by RG flows are often equal. For example if theory ${\cal T}_1$ when deformed by a superpotential leads to theory ${\cal T}_2$, the index of theory ${\cal T}_2$ refined with symmetries visible in the UV is the same as the index of theory ${\cal T}_1$ when one switches off the fugacities corresponding to symmetries broken by the superpotential \cite{Festuccia:2011ws,Rastelli:2016tbz}. Similarly, the index of a theory ${\cal T}_2$ obtained by vev deformation of theory ${\cal T}_1$ is given by a certain residue of the index of ${\cal T}_1$ \cite{Gaiotto:2012xa}. Moreover, two (IR or conformally) dual theories also should have the same index once a proper map between parameters corresponding to global symmetries is found.\footnote{Note that  to match the indices of a pair of dual theories one needs to map correctly the R-symmetry of the two. This is usually done using a-maximization arguments \cite{Intriligator:2003jj}.} 
    
    Our main result is a  version/generalization of this statement. Here we find that certain classes of theories which are related  by RG flows have the same partition function albeit with a non-trivial map between the parameters {\it including the R-symmetries}. The relevant RG flows correspond to fine tuning various physical parameters such that without the fine tuning the theories are massive. For example, exploring the Coulomb branch of the theory the index, as we have discussed, trivializes and is just given by the $\U(1)$ vectors. However, in the examples we have considered the generalized Schur indices of two theories connected by such Coulomb branch deformation, and satisfying conditions we have outlined, are mapped to each other with different values of their $\alpha$ parameters. 

    Our observation is specific to ${\cal N}=2$ SCFTs, and exploits an orders of limits tension between the Schur index and the index obtained turning on Coulomb branch vevs. This tension was phrased in the past as an observation that the Schur index is consistent with mass terms in Lagrangian but nevertheless captures more than just the vacuum. See {\it e.g.}~\cite{DFDZ}. 

    Once we understand that starting from an ${\cal N}=2$ SCFT and computing it's generalized Schur partition function with $\alpha$ we can obtain this partition function for a different theory with $\alpha'(\alpha)$, we can iterate this procedure by, {\it e.g.} adding matter and gauging the symmetry (assuming one can compute the index refined by flavor fugacities, which is true in various cases) with parameter $\alpha''$ to obtain partition functions of new theories. For example, the $(A_3,A_3)$ AD theory can be obtained by a diagonal $\SU(2)$ gauging of two $(A_1,D_4)$ AD theories along with free hypermultiplets \cite{Agarwal:2016pjo,Choi:2017nur}. Another example, is the fact that  the $(A_2,D_4)\simeq \hat{E}_6(\SU(2))$  can be obtained by a diagonal gauging of three copies of $(A_1,A_3)$ theories \cite{Kang:2021lic} . On the other hand  $(A_1,D_4)/(A_1,A_3)$  can be thought of as being obtained in the $\alpha'=\frac\alpha2/\alpha'=\frac\alpha3$ limit of the $\SU(2)$ SQCD. Further gauging (along with hypers when needed) can be thought of as a standard $\alpha=1$ gauging. Thus one can write the partition function as a matrix integral with different fields taken into different powers.\footnote{Let us also mention here that the $\alpha=2$ limit for $(A_3,A_3)$ AD theory gives the Schur index for $[1]-\SU(2)-\SU(3)-[4]$ and $\alpha=3$ limit for $(A_2,D_4)$ AD theory gives the Schur index for $[\SO(8)]-\USp(4)-\SO(4)$ theory. In \cite{Agarwal:2017roi}, an $\mathcal{N}=1$ deformation of the quivers flows to the respective AD theories described. These are a special case of the general AD theories described in \cite{Agarwal:2017roi} and we expect that the limit of the index described here also relates those theories.}

The results of this note open many venues for connecting partition functions of different SCFTs. It will be extremely interesting to find a first principle derivation of the ``experimental'' observations we report here.

\noindent {\bf Acknowledgments:} We would like to thank Leonardo Rastelli, Matteo Sacchi, Wenbin Yan, and Gabi Zafrir for enlightening and useful discussions. SSR is grateful  to the SCGP for hospitality during the program ``Supersymmetric Quantum Field Theories, Vertex Operator Algebras, and Geometry'', where this project was initiated. 
The work of AD is supported in part by NSF grant PHY-2210533 and by the Simons Foundation grant
681267 (Simons Investigator Award). The research of SSR is supported in part by  the Planning and Budgeting committee, by the Israel Science Foundation under grant no. 2159/22, and by BSF-NSF grant no. 2023769.

	\appendix 
	
    \noindent{{\bf TQFT structure of ${\cal Z}(q,\alpha)$}:}
Superconformal indices are invariants of S-duality. In the framework of theories of class ${\cal S}$  \cite{Gaiotto:2009we,Gaiotto:2010okc} this leads to a mathematical statement that the index should be invariant under exchange of puncture symmetries \cite{Gadde:2009kb}. For example, the $\SU(2)$ SQCD has $ \SU(2)^4\subset \SO(8)$ global symmetry with the four $\SU(2)$ factors corresponding to four punctures. S-duality then implies that,
\be 
&&\oint\frac{dz}{4\pi i z} \left(\frac{\Delta(z)(q\, z^{\pm2};q)^2}{(q^{\frac12}z^{\pm1}a^{\pm1}b^{\pm1};q)(q^{\frac12}z^{\pm1}c^{\pm1}d^{\pm1};q)}\right)^\alpha=\;\;\;\;\;\\
&&\;\;\;\oint\frac{dz}{4\pi i z} \left(\frac{\Delta(z)(q\, z^{\pm2};q)^2}{(q^{\frac12}z^{\pm1}a^{\pm1}c^{\pm1};q)(q^{\frac12}z^{\pm1}b^{\pm1}d^{\pm1};q)}\right)^\alpha\,,\nonumber
\ee for any value of $\alpha$. This equality follows mathematically from \cite{FokkovDeBult:2011,Rains:2025fqe} and can be explicitly verified in series expansion.
Moreover, one should be able to expand the index in eigenfunctions \cite{Gaiotto:2012xa} generalizing the Schur polynomials. For example, using the techniques of \cite{Nazzal:2023wtw} one can obtain that the index of $A_1$ class ${\cal S}$ theories on genus $g$ surface with $n$ punctures in the large $g$ limit and $\alpha=2$ as,
\be
{\cal Z}\sim C(g,n)\, \prod_{i=1}^n \psi_0(a_i)\,,
\ee with,
\be 
&&\frac{\psi_0(a)-1}{\left(a-a^{-1}\right)^2}=
\frac{8q}{3}+\frac{\left(135 a^2-74+135a^{-2}\right) q^2}{27}+\\
&&\frac{8  \left(243 a^4-216 a^2+254-216 a^{-2}+243a^{-4}\right) q^3}{243}+\cdots\,.\nonumber
\ee 
Here the wave-function was normalized so that $\psi_0(1)=1$.


\begin{thebibliography}{99}
		
		\bibitem{Gadde:2011ik}
		A.~Gadde, L.~Rastelli, S.~S.~Razamat, W.~Yan,
		``The 4d Superconformal Index from $q$-deformed $2d$ Yang--Mills,''
		Phys. Rev. Lett. \textbf{106}, 241602 (2011)
		[arXiv:1104.3850 [hep-th]].
		
		\bibitem{Gadde:2011uv}
		A.~Gadde, L.~Rastelli, S.~S.~Razamat, W.~Yan,
		``Gauge Theories and Macdonald Polynomials,''
		Commun. Math. Phys. \textbf{319}, 147-193 (2013)
		[arXiv:1110.3740 [hep-th]].



		\bibitem{Kinney:2005ej}
		J.~Kinney, J.~M.~Maldacena, S.~Minwalla and S.~Raju,
		``An Index for 4 dimensional super conformal theories,''
		Commun. Math. Phys. \textbf{275}, 209-254 (2007)
		[arXiv:hep-th/0510251 [hep-th]].

		
		
		
		\bibitem{Beem:2013sza}
		C.~Beem, M.~Lemos, P.~Liendo, W.~Peelaers, L.~Rastelli, B.~C.~van Rees,
		``Infinite Chiral Symmetry in Four Dimensions,''
		Commun. Math. Phys. \textbf{336}, no.3, 1359-1433 (2015)
		[arXiv:1312.5344 [hep-th]].
		
		\bibitem{Cordova:2016uwk}
		C.~Cordova, D.~Gaiotto, S.~H.~Shao, 
		``Infrared Computations of Defect Schur Indices,''
		JHEP \textbf{11}, 106 (2016)
		[arXiv:1606.08429 [hep-th]].
		

		\bibitem{Gaiotto:2024kze}
		D.~Gaiotto and H.~Verlinde,
		``SYK-Schur duality: Double scaled SYK correlators from $N=2$ supersymmetric gauge theory,''
		[arXiv:2409.11551 [hep-th]].

		
		


\bibitem{Gaiotto:2024osr}
D.~Gaiotto and J.~Teschner,
``Schur Quantization and Complex Chern-Simons theory,''
[arXiv:2406.09171 [hep-th]].





\bibitem{Ambrosino:2025qpy}
F.~Ambrosino and D.~Gaiotto,
``Renormalization Group flow in Schur quantization,''
[arXiv:2503.16685 [hep-th]].



		\bibitem{Gaiotto:2024ioj}
		D.~Gaiotto and H.~Kim,
		``3D TFTs from 4d $ \mathcal{N} $ = 2 BPS particles,''
		JHEP \textbf{03}, 173 (2025)
		[arXiv:2409.20393 [hep-th]].


\bibitem{Cordova:2015nma}
C.~Cordova and S.~H.~Shao,
`Schur Indices, BPS Particles, and Argyres-Douglas Theories,''
JHEP \textbf{01}, 040 (2016)
[arXiv:1506.00265 [hep-th]].

\bibitem{Kaidi:2022sng}
J.~Kaidi, M.~Martone, L.~Rastelli and M.~Weaver,
``Needles in a haystack. An algorithmic approach to the classification of 4d $ \mathcal{N} $ = 2 SCFTs,''
JHEP \textbf{03}, 210 (2022)
[arXiv:2202.06959 [hep-th]].


\bibitem{Buican:2024jjl}
M.~Buican,
``Coulomb Branch Operator Algebras and Universal Selection Rules for $\mathcal{N}=2$ SCFTs,''
[arXiv:2406.00178 [hep-th]].


\bibitem{Fredrickson:2017yka}
L.~Fredrickson, D.~Pei, W.~Yan and K.~Ye,
``Argyres-Douglas Theories, Chiral Algebras and Wild Hitchin Characters,''
JHEP \textbf{01}, 150 (2018)
[arXiv:1701.08782 [hep-th]].


\bibitem{Pan:2024hcz}
Y.~Pan and W.~Yan,
``Mirror symmetry for circle compactified 4d $A_1$ class-$S$ theories,''
[arXiv:2410.15695 [hep-th]].


\bibitem{Deb:2025cqr}
A.~Deb, C.~Meneghelli and L.~Rastelli,
``The Nilpotency Index for 4d $\mathcal{N}=2$ SCFTs,''
[arXiv:2503.05975 [hep-th]].



\bibitem{Romelsberger:2005eg}
C.~Romelsberger,
``Counting chiral primaries in N = 1, d=4 superconformal field theories,''
Nucl. Phys. B \textbf{747}, 329-353 (2006)
[arXiv:hep-th/0510060 [hep-th]].

\bibitem{Dolan:2008qi}
F.~A.~Dolan and H.~Osborn,
``Applications of the Superconformal Index for Protected Operators and q-Hypergeometric Identities to N=1 Dual Theories,''
Nucl. Phys. B \textbf{818}, 137-178 (2009)
[arXiv:0801.4947 [hep-th]].

		\bibitem{Gaiotto:2012xa}
		D.~Gaiotto, L.~Rastelli and S.~S.~Razamat,
		``Bootstrapping the superconformal index with surface defects,''
		JHEP \textbf{01}, 022 (2013)
		[arXiv:1207.3577 [hep-th]].


	\bibitem{Beem:2017ooy}
		C.~Beem, L.~Rastelli,
		``Vertex operator algebras, Higgs branches, and modular differential equations,''
		JHEP \textbf{08}, 114 (2018)
		[arXiv:1707.07679 [hep-th]].
        

		\bibitem{Cecotti:2015lab}
		S.~Cecotti, J.~Song, C.~Vafa and W.~Yan,
		``Superconformal Index, BPS Monodromy and Chiral Algebras,''
		JHEP \textbf{11}, 013 (2017)
		[arXiv:1511.01516 [hep-th]].

\bibitem{Cecotti:2010fi}
S.~Cecotti, A.~Neitzke and C.~Vafa,
``R-Twisting and 4d/2d Correspondences,''
[arXiv:1006.3435 [hep-th]].

\bibitem{Kim:2024dxu}
H.~Kim and J.~Song,
``A Family of Vertex Operator Algebras from Argyres-Douglas Theory,''
[arXiv:2412.20015 [hep-th]].


\bibitem{Gadde:2009kb}
		A.~Gadde, E.~Pomoni, L.~Rastelli, S.~S.~Razamat,
		``$S$-duality and $2d$ Topological QFT,''
		JHEP \textbf{03}, 032 (2010)
		[arXiv:0910.2225 [hep-th]].



		\bibitem{Razamat:2014pta}
		S.~S.~Razamat and B.~Willett,
		``Down the rabbit hole with theories of class $ \mathcal{S} $,''
		JHEP \textbf{10}, 099 (2014)
		[arXiv:1403.6107 [hep-th]].

		

		\bibitem{Aharony:2013dha}
		O.~Aharony, S.~S.~Razamat, N.~Seiberg and B.~Willett,
		``3d dualities from 4d dualities,''
		JHEP \textbf{07}, 149 (2013)
		[arXiv:1305.3924 [hep-th]].


\bibitem{Argyres:1995jj}
P.~C.~Argyres and M.~R.~Douglas,
``New phenomena in SU(3) supersymmetric gauge theory,''
Nucl. Phys. B \textbf{448}, 93-126 (1995)
[arXiv:hep-th/9505062 [hep-th]].


\bibitem{Argyres:1995xn}
P.~C.~Argyres, M.~R.~Plesser, N.~Seiberg and E.~Witten,
``New N=2 superconformal field theories in four-dimensions,''
Nucl. Phys. B \textbf{461}, 71-84 (1996)
[arXiv:hep-th/9511154 [hep-th]].


\bibitem{Minahan:1996fg}
J.~A.~Minahan and D.~Nemeschansky,
``An N=2 superconformal fixed point with E(6) global symmetry,''
Nucl. Phys. B \textbf{482}, 142-152 (1996)
[arXiv:hep-th/9608047 [hep-th]].



\bibitem{Minahan:1996cj}
J.~A.~Minahan and D.~Nemeschansky,
``Superconformal fixed points with E(n) global symmetry,''
Nucl. Phys. B \textbf{489}, 24-46 (1997)
[arXiv:hep-th/9610076 [hep-th]].


\bibitem{Argyres:2015ffa}
P.~Argyres, M.~Lotito, Y.~L{\"u} and M.~Martone,
``Geometric constraints on the space of $ \mathcal{N} $ = 2 SCFTs. Part I: physical constraints on relevant deformations,''
JHEP \textbf{02}, 001 (2018)
[arXiv:1505.04814 [hep-th]].

\bibitem{Argyres:2015gha}
P.~C.~Argyres, M.~Lotito, Y.~L{\"u} and M.~Martone,
``Geometric constraints on the space of $ \mathcal{N} $ = 2 SCFTs. Part II: construction of special K{\"a}hler geometries and RG flows,''
JHEP \textbf{02}, 002 (2018)
[arXiv:1601.00011 [hep-th]].


\bibitem{Argyres:2016xmc}
P.~Argyres, M.~Lotito, Y.~L{\"u} and M.~Martone,
``Geometric constraints on the space of $ \mathcal{N}$ = 2 SCFTs. Part III: enhanced Coulomb branches and central charges,''
JHEP \textbf{02}, 003 (2018)
[arXiv:1609.04404 [hep-th]].



\bibitem{Argyres:2020nrr}
P.~Argyres and M.~Martone,
``Construction and classification of Coulomb branch geometries,''
[arXiv:2003.04954 [hep-th]].


	\bibitem{Arakawa:2016hkg}
		T.~Arakawa, K.~Kawasetsu,
		``Quasi-lisse vertex algebras and modular linear differential equations,''
		[arXiv:1610.05865 [math.QA]].
		

		\bibitem{Lee:2023owa}
		K.~Lee, K.~Sun and H.~Wang,
		``On intermediate Lie algebra $E_{7+1/2}$,''
		Lett. Math. Phys. \textbf{114}, no.1, 13 (2024)
		[arXiv:2306.09230 [math-ph]].


\bibitem[Cohen and de~Man(1996)]{cohen1996computational}
Arjeh~M Cohen and Ronald de~Man.
\newblock Computational evidence for deligne's conjecture regarding exceptional
  lie groups.
\newblock \emph{Comptes Rendus de l'Acad{\'e}mie des Sciences. S{\'e}rie 1.
  Math{\'e}matique}, 322\penalty0 (5):\penalty0 427--432, 1996.


\bibitem{Abhishek:2023wzp}
M.~Abhishek, S.~Grover, D.~P.~Jatkar and K.~Singh,
``Looking for the G$_{2}$ Higgs branch of 4D rank 1 SCFTs,''
JHEP \textbf{08}, 026 (2024)
[arXiv:2312.00275 [hep-th]].


\bibitem{Cho:2024civ}
M.~Cho, K.~Maruyoshi, E.~Nardoni and J.~Song,
``Large landscape of 4d superconformal field theories from small gauge theories,''
JHEP \textbf{11}, 010 (2024)
[arXiv:2408.02953 [hep-th]].


\bibitem{Bourget:2020asf}
A.~Bourget, J.~F.~Grimminger, A.~Hanany, M.~Sperling, G.~Zafrir and Z.~Zhong,
``Magnetic quivers for rank 1 theories,''
JHEP \textbf{09}, 189 (2020)
[arXiv:2006.16994 [hep-th]].

\bibitem{Maruyoshi:2016tqk}
K.~Maruyoshi and J.~Song,
``Enhancement of Supersymmetry via Renormalization Group Flow and the Superconformal Index,''
Phys. Rev. Lett. \textbf{118}, no.15, 151602 (2017)
[arXiv:1606.05632 [hep-th]].



\bibitem{Maruyoshi:2016aim}
K.~Maruyoshi and J.~Song,
``$ \mathcal{N}=1 $ deformations and RG flows of $ \mathcal{N}=2 $ SCFTs,''
JHEP \textbf{02}, 075 (2017)
[arXiv:1607.04281 [hep-th]].

        
	
		
	
	





		
        
	


\bibitem{DiPietro:2014bca}
L.~Di Pietro and Z.~Komargodski,
``Cardy formulae for SUSY theories in $d =$ 4 and $d =$ 6,''
JHEP \textbf{12}, 031 (2014)
[arXiv:1407.6061 [hep-th]].



\bibitem{ArabiArdehali:2015ybk}
A.~Arabi Ardehali,
``High-temperature asymptotics of supersymmetric partition functions,''
JHEP \textbf{07}, 025 (2016)
[arXiv:1512.03376 [hep-th]].




\bibitem{Buican:2015ina}
M.~Buican and T.~Nishinaka,
``On the superconformal index of Argyres\textendash{}Douglas theories,''
J. Phys. A \textbf{49}, no.1, 015401 (2016)
[arXiv:1505.05884 [hep-th]].


\bibitem{ArabiArdehali:2023bpq}
A.~Arabi Ardehali, M.~Martone and M.~Rossell\'o,
``High-temperature expansion of the Schur index and modularity,''
[arXiv:2308.09738 [hep-th]].


\bibitem{Bobev:2015kza}
N.~Bobev, M.~Bullimore and H.~C.~Kim,
``Supersymmetric Casimir Energy and the Anomaly Polynomial,''
JHEP \textbf{09}, 142 (2015)
doi:10.1007/JHEP09(2015)142
[arXiv:1507.08553 [hep-th]].

\bibitem{Razamat:2012uv}
S.~S.~Razamat,
``On a modular property of N=2 superconformal theories in four dimensions,''
JHEP \textbf{10}, 191 (2012)
[arXiv:1208.5056 [hep-th]].


\bibitem{Shapere:2008zf}
A.~D.~Shapere and Y.~Tachikawa,
``Central charges of N=2 superconformal field theories in four dimensions,''
JHEP \textbf{09}, 109 (2008)
[arXiv:0804.1957 [hep-th]].


\bibitem{Agarwal:2016pjo}
 P.~Agarwal, K.~Maruyoshi and J.~Song,
 ``$ \mathcal{N} $ =1 Deformations and RG flows of $ \mathcal{N} $ =2 SCFTs, part II: non-principal deformations,''
 JHEP \textbf{12}, 103 (2016)
 [arXiv:1610.05311 [hep-th]].


\bibitem{Martone:2021ixp}
M.~Martone,
``Testing our understanding of SCFTs: a catalogue of rank-2 $ \mathcal{N} $ = 2 theories in four dimensions,''
JHEP \textbf{07}, 123 (2022)
[arXiv:2102.02443 [hep-th]].



\bibitem{Martone:2021drm}
M.~Martone and G.~Zafrir,
``On the compactification of 5d theories to 4d,''
JHEP \textbf{08}, 017 (2021)
[arXiv:2106.00686 [hep-th]].



\bibitem{Festuccia:2011ws}
G.~Festuccia and N.~Seiberg,
``Rigid Supersymmetric Theories in Curved Superspace,''
JHEP \textbf{06}, 114 (2011)
[arXiv:1105.0689 [hep-th]].


		\bibitem{Rastelli:2016tbz}
		L.~Rastelli and S.~S.~Razamat,
		``The supersymmetric index in four dimensions,''
		J. Phys. A \textbf{50}, no.44, 443013 (2017)
		[arXiv:1608.02965 [hep-th]].

        


\bibitem{Intriligator:2003jj}
K.~A.~Intriligator and B.~Wecht,
``The Exact superconformal R symmetry maximizes a,''
Nucl. Phys. B \textbf{667}, 183-200 (2003)
[arXiv:hep-th/0304128 [hep-th]].

		
		\bibitem{DFDZ}
		T.~Dumitrescu, G.~Festuccia, M.~Del~Zotto,
		\href{https://www.youtube.com/watch?v=ib4v9lNPKS0&t=27s}{Talk at NatiFest 2016}

 \bibitem{Choi:2017nur}
 J.~Choi and T.~Nishinaka,
 ``On the chiral algebra of Argyres-Douglas theories and S-duality,''
 JHEP \textbf{04}, 004 (2018)
 [arXiv:1711.07941 [hep-th]].

\bibitem{Kang:2021lic}
 M.~J.~Kang, C.~Lawrie and J.~Song,
 ``Infinitely many 4D N=2 SCFTs with a=c and beyond,''
 Phys. Rev. D \textbf{104}, no.10, 105005 (2021)
 [arXiv:2106.12579 [hep-th]].



 \bibitem{Agarwal:2017roi}
 P.~Agarwal, A.~Sciarappa and J.~Song,
 ``$ \mathcal{N} $ =1 Lagrangians for generalized Argyres-Douglas theories,''
 JHEP \textbf{10}, 211 (2017)
 [arXiv:1707.04751 [hep-th]].

 

	\bibitem{Gaiotto:2009we}
		D.~Gaiotto,
		``$\mathcal{N}=2$ dualities,''
		JHEP \textbf{08}, 034 (2012)
		[arXiv:0904.2715 [hep-th]].

  
\bibitem{Gaiotto:2010okc}
D.~Gaiotto, G.~W.~Moore and A.~Neitzke,
``Four-dimensional wall-crossing via three-dimensional field theory,''
Commun. Math. Phys. \textbf{299}, 163-224 (2010)
[arXiv:0807.4723 [hep-th]].


\bibitem{FokkovDeBult:2011}
F.~J. van~de~Bult, ``An elliptic hypergeometric integral with $W(F_4)$ symmetry,'', Ramanujan J. {\bf 25} (2011), no.~1, 1--20
[arXiv:0909.4793[math]].


\bibitem{Rains:2025fqe}
E.~Rains and H.~Rosengren,
``Q-operators for the Ruijsenaars model,''
[arXiv:2503.18057 [math-ph]].



\bibitem{Nazzal:2023wtw}
B.~Nazzal, A.~Nedelin and S.~S.~Razamat,
``Ground state wavefunctions of elliptic relativistic integrable Hamiltonians,''
Nucl. Phys. B \textbf{996}, 116364 (2023)
[arXiv:2305.09718 [hep-th]].

\end{thebibliography}
\end{document}